\documentclass[aps,showpacs,nofootinbib,twocolumn,superscriptaddress]{revtex4}
\usepackage{amsmath}
\usepackage{amsthm}
\usepackage{mathtools}
\usepackage{amssymb}
\usepackage{subfigure}
\usepackage[english]{babel}
\begin{document}

\title{\large\bf Experimental investigation of multi-observable uncertainty relations}

\affiliation{School of Physical Sciences, University of Chinese
Academy of Sciences, \\ YuQuan Road 19A, Beijing 100049, China}

\affiliation{College of Materials Science and Opto-Electronic
Technology, \\ University of Chinese Academy of Sciences, YuQuan Road 19A, Beijing 100049, China}

\affiliation{Department of Physics \& Astronomy, York University, Toronto, Canada ON M3J 1P3}

\affiliation{Key Laboratory of Vacuum Physics, University of Chinese Academy of Sciences, \\ YuQuan Road 19A, Beijing 100049, China}

\author{Zhi-Xin Chen}

\affiliation{School of Physical Sciences, University of Chinese
Academy of Sciences, \\ YuQuan Road 19A, Beijing 100049, China}

\author{Jun-Li Li}

\affiliation{College of Materials Science and Opto-Electronic
Technology, \\ University of Chinese Academy of Sciences, YuQuan Road 19A, Beijing 100049, China}

\affiliation{Key Laboratory of Vacuum Physics, University of Chinese Academy of Sciences, \\ YuQuan Road 19A, Beijing 100049, China}

\author{Qiu-Cheng Song}

\affiliation{School of Physical Sciences, University of Chinese
Academy of Sciences, \\ YuQuan Road 19A, Beijing 100049, China}

\author{Hui Wang}

\affiliation{College of Materials Science and Opto-Electronic
Technology, \\ University of Chinese Academy of Sciences, YuQuan Road 19A, Beijing 100049, China}

\author{S. M. Zangi}

\affiliation{School of Physical Sciences, University of Chinese
Academy of Sciences, \\ YuQuan Road 19A, Beijing 100049, China}

\author{Cong-Feng Qiao}

\email{qiaocf@ucas.ac.cn}

\affiliation{School of Physical Sciences, University of Chinese
Academy of Sciences, \\ YuQuan Road 19A, Beijing 100049, China}

\affiliation{Department of Physics \& Astronomy, York University, Toronto, Canada ON M3J 1P3}

\affiliation{Key Laboratory of Vacuum Physics, University of Chinese Academy of Sciences, \\ YuQuan Road 19A, Beijing 100049, China}

\begin{abstract}
The uncertainty relation is a distinguishing feature of quantum theory, characterizing the incompatibility 
of noncommuting observables in the preparation of  quantum states. Recently, many uncertainty relations 
were proposed with improved lower bounds and were deemed capable of incorporating multiple observables. Here 
we report an experimental verification of seven uncertainty relations of this type with single-photon 
measurements. The results, while confirming these uncertainty relations, show as well the relative stringency 
of various uncertainty lower bounds.
\end{abstract}
\maketitle

\section{introduction}

The uncertainty relation lies at the heart of quantum theory,  and plays an important role in quantum information 
sciences. For example, it is crucial in the studies of quantum cryptography \cite{Fuchs,Renes}, entanglement detections \cite{Hofmann,OGuhne}, quantum steering \cite{Schneeloch,YZZhen}, quantum nonlocalities \cite{Oppenheim,ZhihAhnJia}, and so on. The original concept of the uncertainty relation was introduced by Heisenberg while demonstrating the impossibility 
of the simultaneous precise measurement of the position and momentum of an electron \cite{heis}. Later, Kennard \cite{Kennard}, Weyl \cite{Weyl},  Robertson \cite{Robertson}, and Schr\"odinger \cite{schrodinger} derived mathematically rigorous uncertainty relations, among which the most famous one is the Heisenberg-Robertson uncertainty relation
\begin{eqnarray}\label{Robertson}
\Delta A^2\Delta B^2\geq
 \left|  \frac{1}{2} \langle\psi|[A,B]|\psi\rangle\right|^2 \; . \label{Robertson1}
\end{eqnarray}
Here $\Delta A^2$ and $\Delta B^2$ are variances of observables $A$ and $B$. The uncertainty relation (\ref{Robertson1}) states that the product of the variances of the measurement results of the incompatible observables is lower bounded by 
the expectation value of the commutator $\langle[A,B]\rangle$, which may be zero even when neither of the two variances is.

To get a better lower bound, Maccone and Pati proposed the following uncertainty relations \cite{mp}:
\begin{align}
\Delta A^2 + \Delta B^2 & \geq
\pm i\langle\psi|[A,B]|\psi\rangle+|\langle\psi|A \pm iB|\psi^\perp\rangle|^2, \label{Maccone} \\
\Delta A^2 + \Delta B^2 & \geq
\frac 12|\langle\psi^\perp_{A+B}|A+B|\psi\rangle|^2\; , \label{Maccone2p}
\end{align}
where $\langle \psi|\psi^\perp\rangle = 0$, $|\psi^\perp_{A+B}\rangle\propto(A+B-\langle A + B\rangle)|\psi\rangle$, 
and the sign on the right-hand side of the inequality (\ref{Maccone}) takes $+(-)$ when $ i\langle\psi|[A,B]|\psi\rangle$ 
is positive (negative). After these, more state-dependent \cite{bannur,sun,song,Debasis,xiao,zhang} and state-independent  \cite{Huang, L1, L2, Branciard, Schwonnek, YunlongXiao} lower bounds of uncertainty relations were also obtained. 
The generalization and improvement mainly focused on the uncertainty relation capable of dealing with more than two incompatible observables, i.e., the Heisenberg-type uncertainty relation for three canonical observables \cite {Kechrimparis}, uncertainty relations for angular momentum \cite{dammeier}, and  arbitrary incompatible observables \cite{song}. There are also uncertainty relations with even more incompatible observables formulated in products \cite{qin,naihuan} or sums \cite{long,chenfei,songqc} of variances.

Equations (\ref{Maccone}) and (\ref{Maccone2p}) were tested in \cite{xue}, and an uncertainty relation of triple spin components 
of qubit systems was carried out in \cite{dufei}. Though there is a growing number of uncertainty relations involving multiple incompatible observables, fewer experiments have been carried out in testing their validity \cite{dufei}. 
Meanwhile, as these uncertainty relations differ by lower bounds, an experimental investigation not only provides 
a verification, but also a ranking of the uncertainty relations over their lower bounds.

In this paper, we report a systematic experimental investigation of uncertainty relations, where three uncertainty 
relations for triple observables and four for $N$ observables are tested with single-photon measurements. With the qubit-system experiments, the results verify the validity of these uncertainty relations, and the superiority 
of our proposed uncertainty relations \cite{song,songqc} is also exhibited.

\section{Theoretical framework}

The uncertainty relation, as one of the most fundamental elements of quantum mechanics, captures the incompatibility of noncommuting observables. The conventional Robertson-type uncertainty relations concern only two observables. However, in practice, more incompatible observables may appear in the measurement, e.g., angular momentum has three noncommuting components. Hence, it is important to study uncertainty relations for multiple incompatible observables. 
Although uncertainty relations for three canonical observables  \cite {Kechrimparis} and angular momentum \cite{dammeier} exist, our discussed ones are universally valid for arbitrary observables.

\subsection{Uncertainty relations for triple and $N$ observables}

In \cite{song}, two uncertainty relations for triple observables were proposed. The first one reads
\begin{align}\label{our1}
&\Delta A^2+\Delta B^2+\Delta C^2\geq\notag \\
&\frac{1}{3}\Delta(A+B+C)^2 + \frac{\sqrt{3}}{3}|\langle[A,B,C]\rangle|\; ,
\end{align}
where  $\langle[A,B,C]\rangle\equiv\langle[A,B]\rangle+\langle[B,C]\rangle+\langle[C,A]\rangle$,
and the second relation reads \cite{song}
\begin{align}\label{our2}
&\Delta A^2+\Delta B^2+\Delta C^2\geq\notag\\
&\frac{\sqrt{3}}{3}\{|\langle[A,B]\rangle|+|\langle[B,C]\rangle|+|\langle[C,A]\rangle|\} \; .
\end{align}
Equation (\ref{our2}) is definitely stronger than the generalization of the Heisenberg-Robertson uncertainty relation, 
i.e., a  simple addition of three pairwise uncertainty relations as the following:
\begin{align} \label{heisenberg2}
&\Delta A^2 + \Delta B^2 + \Delta C^2 \geq  \Delta A\Delta B + \Delta B \Delta C + \Delta C\Delta A \nonumber \\
&\geq\frac{1}{2}\{|\langle[A,B]\rangle|+|\langle[B,C]\rangle|+|\langle[C,A]\rangle|\}.
\end{align}
Note that Eq. (\ref{our2}) has been verified using single spin measurements in diamond afterwards \cite{dufei}.

For the case of $N$ incompatible observables, one has the following two inequalities from inequality (\ref{Maccone2p}):
\begin{eqnarray}
\sum_{i=1}^{N}(\Delta A_{i})^{2}\geq\frac{1}{2(N-1)}\sum_{1\leq i<j\leq N}[\Delta(A_{i}+A_{j})]^{2},\label{Maccone3p}\\
\sum_{i=1}^{N}(\Delta A_{i})^{2}\geq\frac{1}{2(N-1)}\sum_{1\leq i<j\leq N}[\Delta(A_{i}-A_{j})]^{2}.\label{Maccone3m}
\end{eqnarray}
But these two relations can be further improved. A stronger uncertainty relation for $N$ observables was derived by 
Chen and Fei \cite{chenfei}, i.e.,
\begin{align}\label{chen}
&\sum_{i=1}^{N}(\Delta A_{i})^{2}
\geq\frac{1}{N-2}\sum_{1\leq i<j\leq N}\left[\Delta (A_{i}+A_{j})\right]^2\notag\\
&-\frac{1}{(N-1)^2(N-2)}\left[\sum_{1\leq i<j\leq N}\Delta (A_{i}+A_{j})\right]^2,
\end{align}
which has a tighter lower bound than the relation (\ref{Maccone3p}) \cite{chenfei}, but it is not always tighter 
than the inequality (\ref{Maccone3m}) \cite{songqc}. In \cite{songqc} a stronger uncertainty relation for $N$ incompatible observables was found, i.e.,
\begin{align}\label{our3}
&\sum_{i=1}^{N}(\Delta A_{i})^{2}
\geq\frac{1}{N}\left[\Delta\left(\sum_{i=1}^{N}A_{i}\right)\right]^{2}\notag\\
&+\frac{2}{N^2(N-1)}\left[\sum_{1\leq i<j\leq N}\Delta(A_{i}-A_{j})\right]^{2},
\end{align}
which has a more stringent bound than the uncertainty relations
Eqs. (\ref{Maccone3p}), (\ref{Maccone3m}), and (\ref{chen}) in qubit systems.

\subsection{Measurements in qubit systems}

Here we choose the following three Pauli operators as three incompatible observables: 
\begin{align}
A&=\sigma_{x}=\begin{pmatrix}
0 & 1\\1 & 0
\end{pmatrix},~~~\\
B&=\sigma_{y}=\begin{pmatrix}
0 & -i \\i & 0
\end{pmatrix},~~~\\
C&=\sigma_{z}=\begin{pmatrix}
1 & 0 \\0 & -1
\end{pmatrix},~~~
\end{align}
and a qubit state parameterized by $\theta$ and $\phi$ as
\begin{align}\label{state}
|\psi(\theta,\phi)\rangle=\cos\frac{\theta}{2}|0\rangle+e^{i \phi}\sin\frac{\theta}{2}|1\rangle ,
\end{align}
where $|0\rangle$ and $|1\rangle$ are eigenstates of $\sigma_{z}$ corresponding to the eigenvalues $1$ and $-1$, respectively,  $\theta\in[0,\pi]$ and $\phi\in[0,2\pi]$.

The variances $(\Delta \sigma_{i})^2=1-\langle \sigma_{i}\rangle^{2}, (i=x,y,z)$ can be computed by the measured 
expectation values of $\sigma_{i}$. All the terms on the left-hand side (LHS) of relations (\ref{our1}), (\ref{our2}), 
and (\ref{heisenberg2}) can be rewritten as $3-V$, where $V=\langle \sigma_{x}\rangle^{2}+\langle \sigma_{y}\rangle^{2}+\langle \sigma_{z}\rangle^{2}$. The right-hand side (RHS) of relation (\ref{our1})
can be rewritten as $(3-V-2D)/3+2 E/\sqrt{3}$, where
$D=\langle \sigma_x\rangle\langle \sigma_y\rangle
+\langle \sigma_y\rangle\langle \sigma_z\rangle
+\langle \sigma_z\rangle\langle \sigma_x\rangle$
and
$E=|\langle \sigma_{x}\rangle+\langle \sigma_{y}\rangle+\langle \sigma_{z}\rangle|$.
Similarly, we can rewrite the terms on the RHS of relations (\ref{our2}) and (\ref{heisenberg2}) as $2 H/\sqrt{3}$ 
and $H$, respectively, where $H=|\langle \sigma_{x}\rangle|+|\langle \sigma_{y}\rangle|+|\langle \sigma_{z}\rangle|$.
Hence, the relations (\ref{our1}),  (\ref{our2}), and  (\ref{heisenberg2}) can be expressed as
\begin{align}
3-V&\geq\frac{1}{3}(3-V-2D)+\frac{2\sqrt{3}}{3} E\; , \label{uncertainty-I}\\
3-V&\geq  \frac{2\sqrt{3}}{3} H\; , \label{uncertainty-II}\\
3-V&\geq  H\; . \label{uncertainty-III}
\end{align}

All the terms on the LHS of relations (\ref{Maccone3p}), (\ref{Maccone3m}), (\ref{chen}), and (\ref{our3})
can be rewritten as $3-V$. The terms on the RHS of inequalities (\ref{Maccone3p}) and (\ref{Maccone3m})
can be rewritten as $(3-V-D)/2$ and $(3-V+D)/2$, respectively. Hence, inequalities (\ref{Maccone3p}) and 
(\ref{Maccone3m}) can be expressed as
\begin{align}
3-V&\geq\frac{1}{2}(3-V-D)\; , \label{uncertainty-IV}\\
3-V&\geq\frac{1}{2}(3-V+D)\; . \label{uncertainty-V}
\end{align}
Similarly we can write relations (\ref{chen}) and (\ref{our3}) in the following forms:
\begin{align}
3-V&\geq2(3-V-D)-\frac{1}{4}(L_{+}+M_{+}+N_{+})^2\; , \label{uncertainty-VI}\\
3-V&\geq\frac{1}{3}(3-V-2D)+\frac{1}{9}(L_{-}+M_{-}+N_{-})^2\; , \label{uncertainty-VII}
\end{align}
where
\begin{align}
L_{\pm}=&\sqrt{2-(\langle \sigma_x\rangle \pm \langle \sigma_y\rangle)^2}\; ,\\
M_{\pm}=&\sqrt{2-(\langle \sigma_y\rangle \pm \langle \sigma_z\rangle)^2}\; ,\\
N_{\pm}=&\sqrt{2-(\langle \sigma_z\rangle \pm \langle \sigma_x\rangle)^2}\; .
\end{align}
In this way, the quantities in Eqs. (\ref{uncertainty-I}) to (\ref{uncertainty-VII}) constitute all the measurements that we need for the experimental verification of these seven relations in qubit systems.

\section{Experimental implementation}

\begin{figure}\centering
\includegraphics[width=0.45\textwidth]{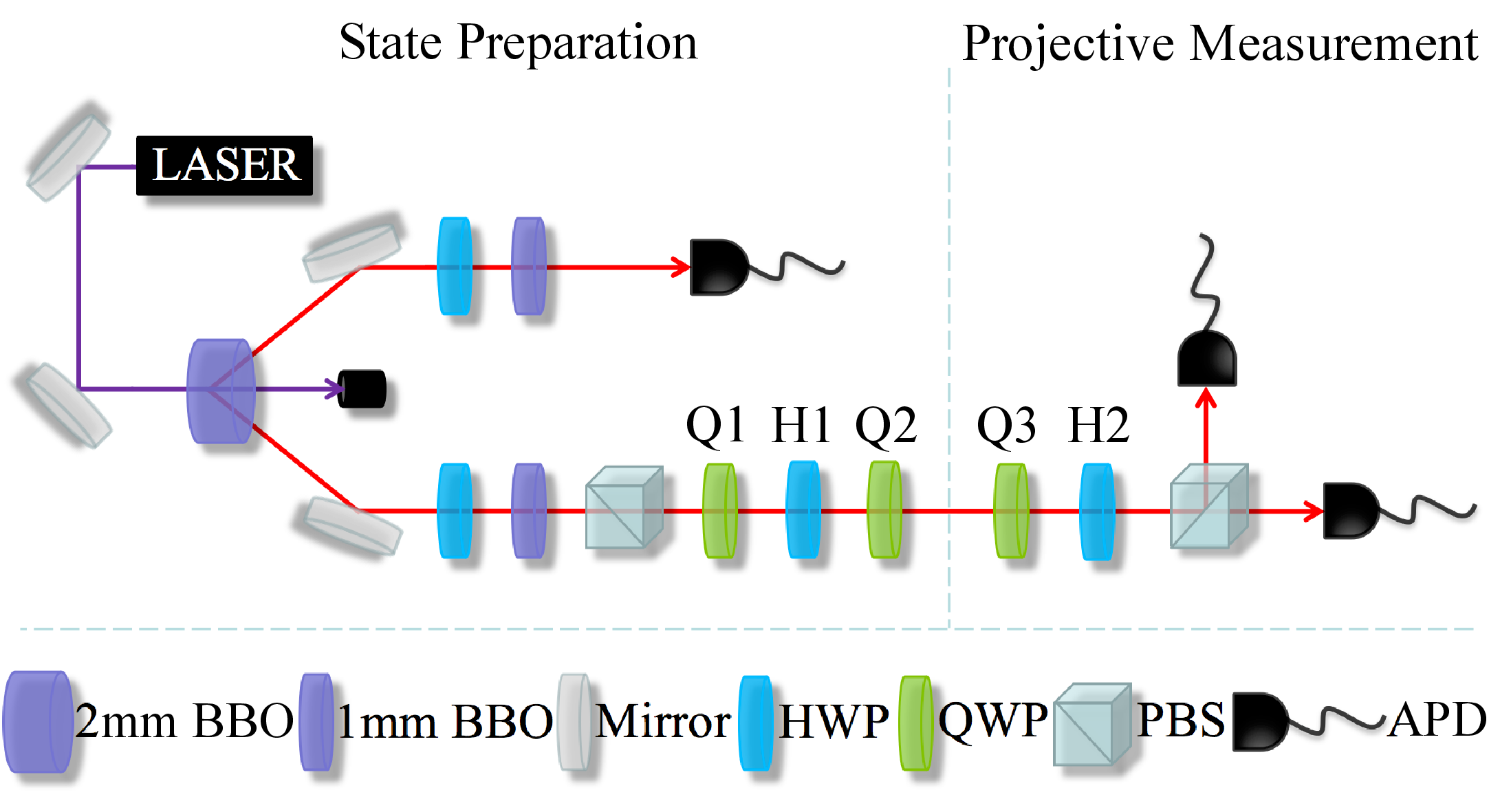}
\caption{Experimental setup. In the state preparation stage, a photon pair is produced via type-II spontaneous 
parametric down-conversion (SPDC) in a BBO crystal. Half-wave plate (HWP) and 1-mm BBO in each path are used to 
compensate the birefringent walk-off effect in the 2-mm BBO. Polarizing beam splitter (PBS), quarter wave plate 
(QWP, Q1), H1, and Q2 are used to generate qubit states. In the projective measurement stage, Q3, H2, and PBS are used 
for measuring the expectation values of the observables.}\label{f1}
\end{figure}
\vspace{.1cm}

\begin{figure} \centering
\includegraphics[width=0.4\textwidth]{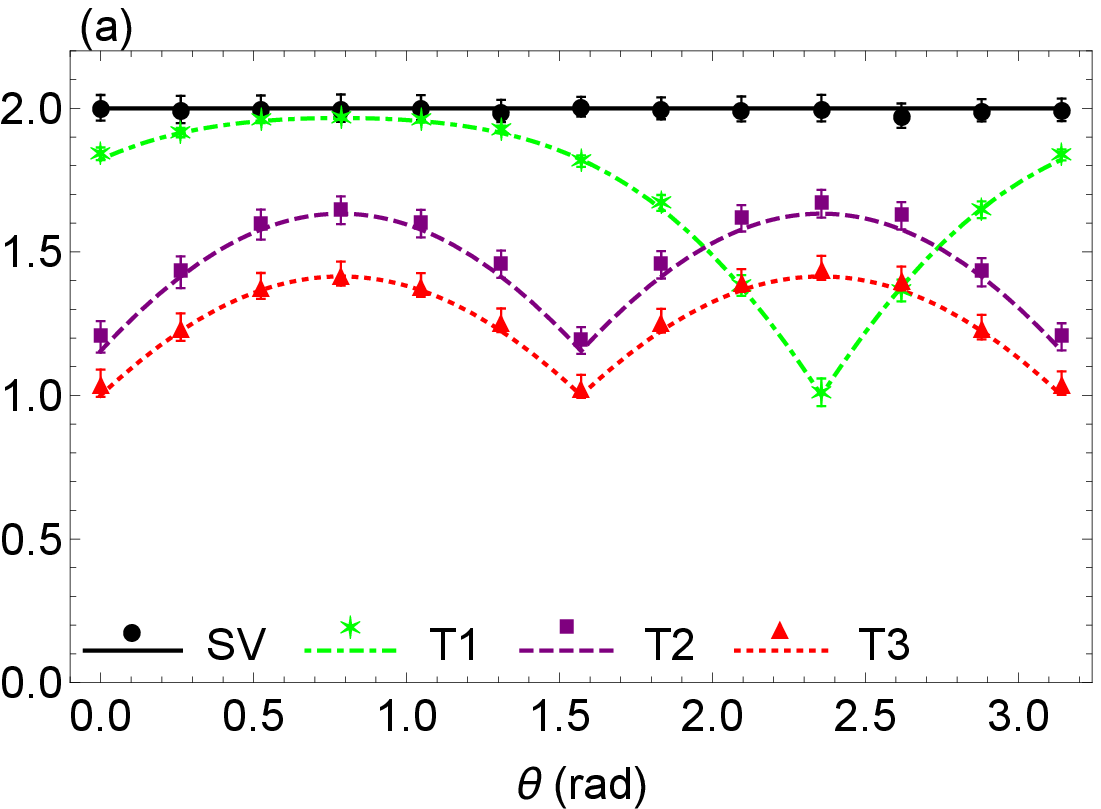}
\includegraphics[width=0.4\textwidth]{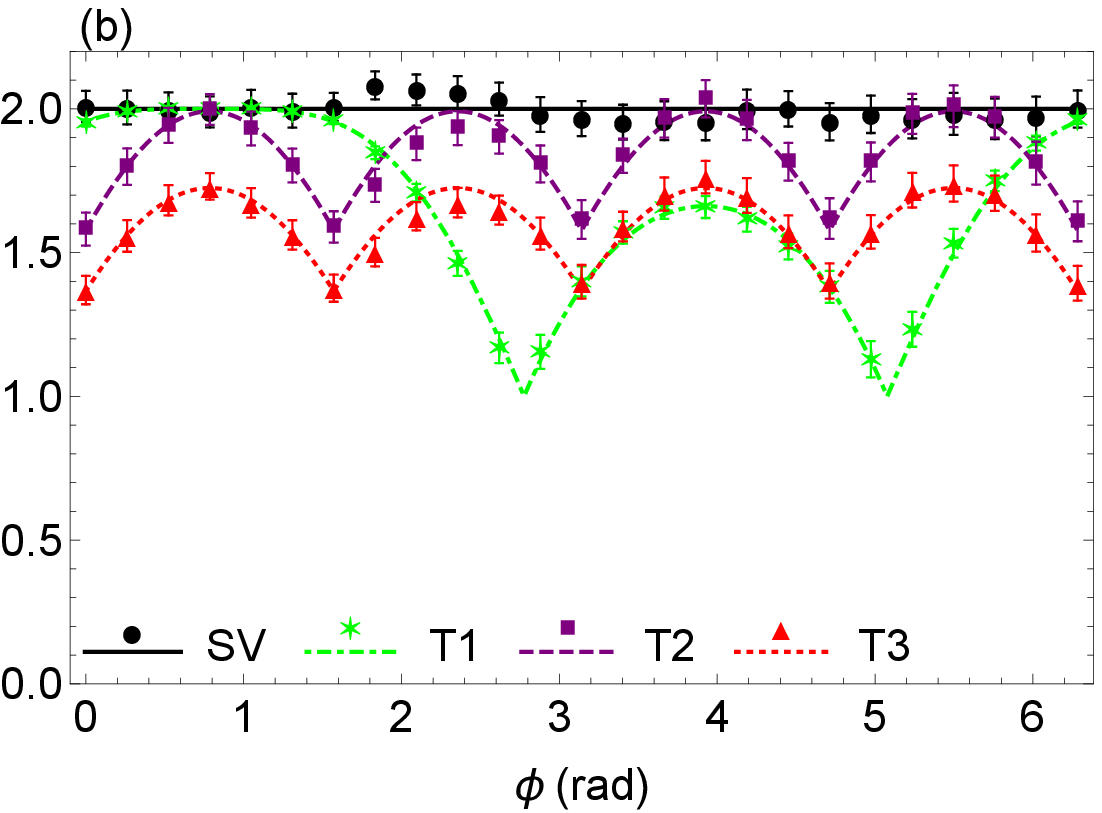}
\caption{Experimental results of uncertainty relations for three observables. (a) Experimental results for $A=\sigma_x$, $B=\sigma_y$, and $C=\sigma_z$, with a family of states $|\psi(\theta,0)\rangle = \cos(\theta/2) |0\rangle + \sin(\theta/2)|1\rangle$. (b) Experimental results for $\sigma_x$, $\sigma_y$, and $\sigma_z$, with states $|\psi(\pi/3,\phi)\rangle = \sqrt{3}/{2} |0\rangle + e^{i\phi}/2|1\rangle$. In (a) and (b), 
the solid black line corresponds to the LHS of inequalities (\ref{our1}), (\ref{our2}), and  (\ref{heisenberg2}), i.e., the sum of the variances (SV) $(\Delta \sigma_{x})^{2}+(\Delta \sigma_{y})^{2}+(\Delta \sigma_{z})^{2}$. The black circles represent the measured SV. The dot-dashed green, 
dashed purple, and dotted red curves, in turn, represent the theoretical values of T1, T2, and T3, where T1, T2, 
and T3 denote the RHS of relations (\ref{our1}), (\ref{our2}), and  (\ref{heisenberg2}), respectively. The green 
stars, purple squares, and red triangles, in turn, represent the experimental values of T1, T2, and T3. Error bars 
represent $\pm1$ standard deviation.
}\label{f2}
\end{figure}
\vspace{.2cm}

\begin{figure}\centering
\includegraphics[width=0.4\textwidth]{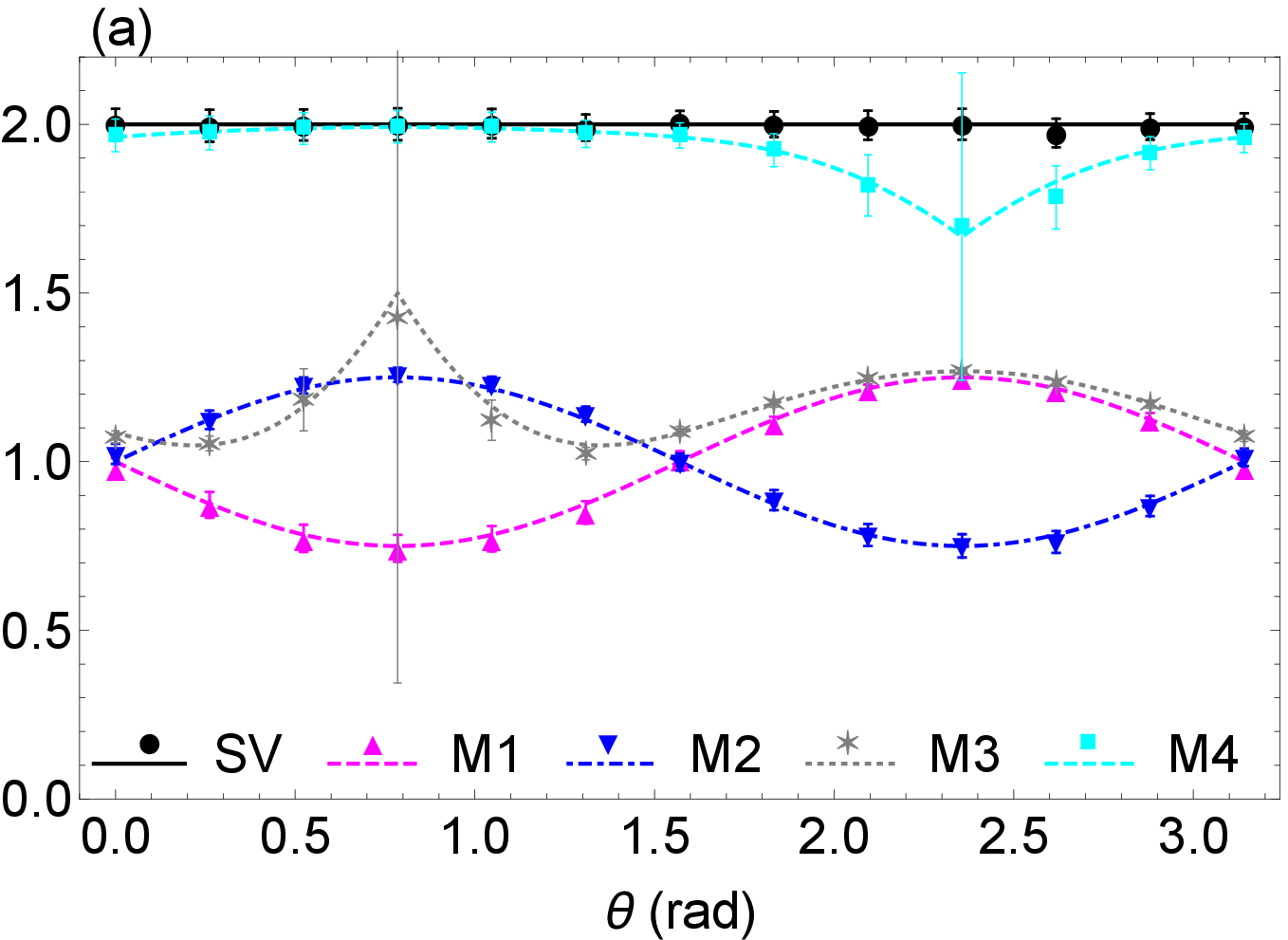}
\includegraphics[width=0.4\textwidth]{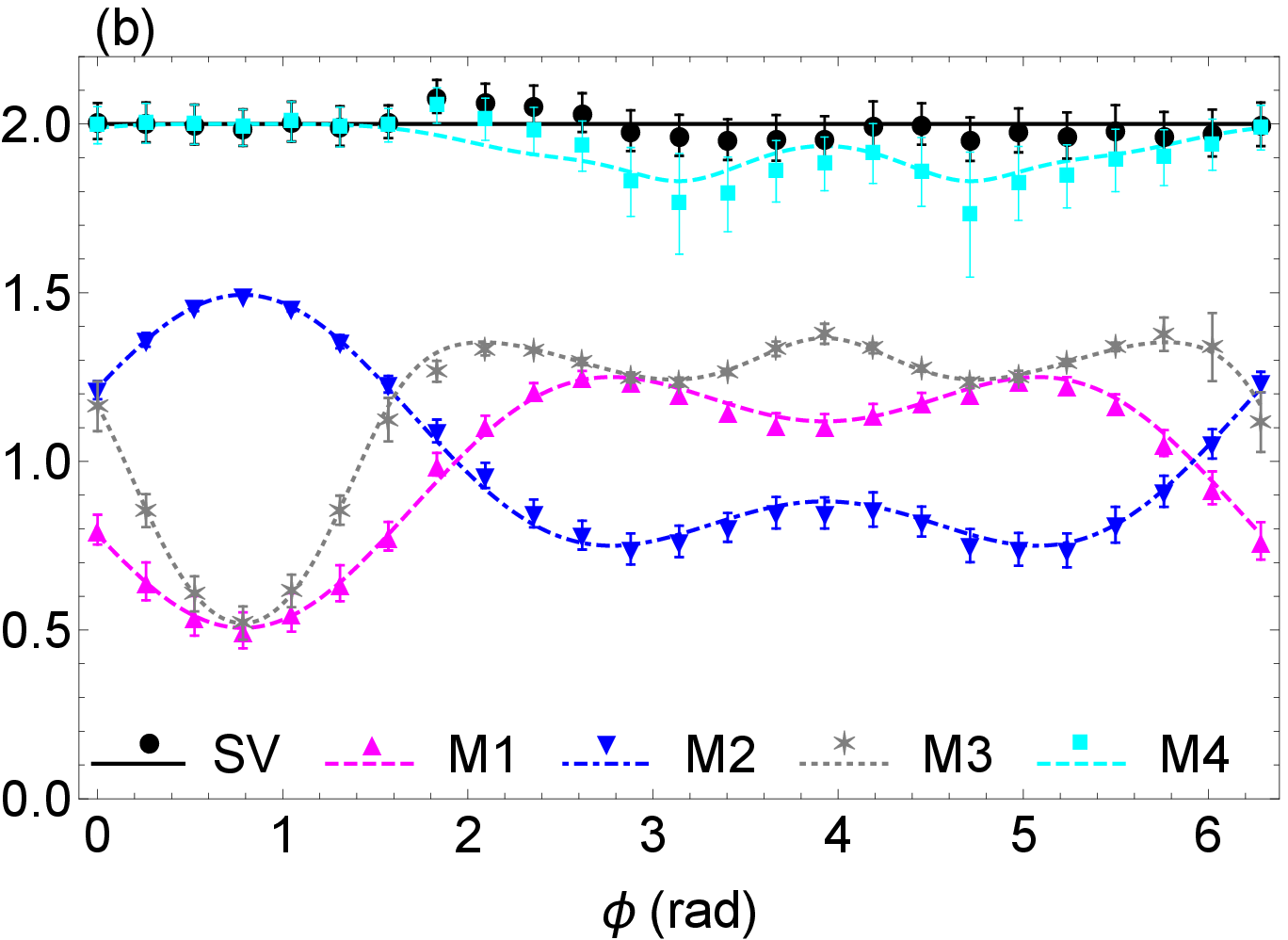}
\caption{Experimental results of uncertainty relations for $N$ observables ($N=3$).
(a) Experimental results for $A=\sigma_x$, $B=\sigma_y$, and $C=\sigma_z$, with a family of states $|\psi(\theta,0)\rangle = \cos(\theta/2) |0\rangle + \sin(\theta/2)|1\rangle$. (b) Experimental results for $\sigma_x$, $\sigma_y$, and $\sigma_z$, with states $|\psi(\pi/3,\phi)\rangle = \sqrt{3}/{2} |0\rangle + e^{i\phi}/2|1\rangle$. In (a) and (b), the solid black line corresponds to the LHS of inequalities (\ref{Maccone3p}), (\ref{Maccone3m}), (\ref{chen}), and (\ref{our3}), i.e., the sum of the variances (SV) $(\Delta \sigma_{x})^{2}+(\Delta \sigma_{y})^{2}+(\Delta \sigma_{z})^{2}$. The black circles represent the measured SV. The dashed magenta, dot-dashed 
blue, dotted gray, and dashed cyan curves, in turn, represent the theoretical values of M1, M2, M3, and M4, where M1, 
M2, M3, and M4 denote the RHS of relations (\ref{Maccone3p}), (\ref{Maccone3m}), (\ref{chen}), and (\ref{our3}), 
respectively. The magenta up-triangles, blue down-triangles, gray stars, and cyan squares, in turn, represent the 
experimental values of M1, M2, M3, and M4.
}\label{f3}
\end{figure}
\vspace{.2cm}

We experimentally verify and compare the uncertainty relations of Eqs. (\ref{our1}) to (\ref{our3}) to single-photon measurements. Figure \ref{f1} shows the experimental setup.  There are two stages, the preparation and projective measurement 
of quantum states. In the preparation stage, photon pairs with wavelength 810 nm are generated via type-\uppercase\expandafter{\romannumeral2}  spontaneous parametric down-conversion (SPDC) in a 2-mm-thick 
nonlinear-barium borate (BBO) crystal, pumped by a 405-nm continuous wave diode laser  with 77 mW of power. We use a half-wave plate (HWP) and a 1-mm BBO in each path to compensate the birefringent walk-off effect in the 2-mm BBO. The trigger photon in the upper path and the signal photon in the lower path are simultaneously detected by two 
single-photon avalanche photodiodes (APD) which are connected with the coincidence counter. The rate of coincidence count is about  2400 $s^{-1}$.

We prepare a qubit in two series of quantum states and use the horizontal and vertical polarizations of the photon to 
encode the basis states  $|0\rangle$ and $|1\rangle$, respectively. The signal photon is initialized to the state 
$|0\rangle$ by a polarizing beam splitter (PBS) then passes through a quarter wave plate (QWP, Q1), a half-wave 
plate (HWP, H1), and Q2 in the state preparation stage. By setting angles of the Q1, H1, and Q2, we can get the 
Eq. (\ref{state}) qubit state $|\psi(\theta,\phi)\rangle = \cos({\theta}/{2}) |0\rangle + e^{i\phi}\sin({\theta}/{2})|1\rangle$  under two different conditions: (i) For $\phi=0$, we choose 
$\theta=n \pi/12$ $(n =0, 1, . . . ,12)$, i.e., a total of 13 states, (ii) For $\theta=\pi/3$, we choose 
$\phi=n\pi/12$  $(n =0, 1, . . . ,24)$, i.e., a total of 25 states. In the measurement stage, Q3, H2, and PBS 
are used to realize the projective measurement of the observables $\sigma_{x}$, $\sigma_{y}$, and $\sigma_{z}$ under 
some certain angles of Q3 and H2.

In Fig. \ref{f2}, we show the experimental verifications of the uncertainty relations (\ref{our1}), (\ref{our2}), and (\ref{heisenberg2}).
We use the states $|\psi(\theta,0)\rangle  = \cos({\theta}/{2}) |0\rangle + \sin({\theta}/{2})|1\rangle$ and $|\psi(\frac{\pi}{3},\phi)\rangle = {\sqrt{3}}/{2} |0\rangle + e^{i\phi}/2|1\rangle$  in Figs. \ref{f2}(a) and \ref{f2}(b), respectively. Our experimental results fit the theoretical predictions well. By comparison, all experimental data of our uncertainty relation (\ref{our2}) are above the generalization of the Heisenberg-Robertson uncertainty relation (\ref{heisenberg2}), which means that our inequality (\ref{our2}) is better than relation (\ref{heisenberg2}), while our other relation (\ref{our1}) is better in some areas.

Similarly, in Fig. \ref{f3}, we plot the experimental results of uncertainty relations (\ref{Maccone3p}), (\ref{Maccone3m}), (\ref{chen}), and (\ref{our3}) for $N=3$ using states $|\psi(\theta,0)\rangle$ and $|\psi(\frac{\pi}{3}, \phi)\rangle$ in Figs. \ref{f3}(a) and \ref{f3}(b), respectively.  Our experimental results fit the theoretical predictions well. The results also show our uncertainty relation (\ref{our3}) is much stronger than relations (\ref{Maccone3p}), 
(\ref{Maccone3m}), and (\ref{chen}) in qubit systems.

The LHS of all inequalities [(\ref{our1}), (\ref{our2}), and (\ref{heisenberg2}) and (\ref{Maccone3p}), (\ref{Maccone3m}), (\ref{chen}), and (\ref{our3})] is the sum of the variances of  three Pauli operators, which is constant for the family of 
states (\ref{state}) and equal to $2$. The experimental results for $(\Delta \sigma_{x})^{2}+(\Delta \sigma_{y})^{2} + (\Delta \sigma_{z})^{2}$ fit the theoretical predictions well.
All the RHS of these seven inequalities can be compared together, including the uncertainty relations for $N$ observables ($N=3$). We find that the uncertainty relations (\ref{our1}), (\ref{our2}), and (\ref{our3}) usually have 
more stringent lower bounds than others as shown in Figs. \ref{f2} and \ref{f3}.

The experimental errors can occur due to the fluctuation of photon counts and the imperfection of wave plates. 
For the same quantum state, even though the original data errors are identical, the final errors of lower bounds of different inequalities are not equal because the errors' propagation of different functions are different. For instance, 
in Fig. \ref{f3}(a) the error bars of the lower bounds of relations (\ref{chen}) at $\theta\simeq \pi/4$ and (\ref{our3}) 
at $\theta\simeq 3\pi/4$ are longer than the others because their error transfer coefficients at the corresponding 
point are much larger.

Note that the lower bounds plotted in Figs. \ref{f2} and \ref{f3} are all state-dependent, so they can be expressed in terms of Bloch parameters of the quanum  state in (\ref{state}).  Since $\langle \sigma_{x} \rangle = \sin\theta \cos\phi $, $\langle \sigma_{y} \rangle = \sin\theta \sin\phi $, and $\langle \sigma_{z} \rangle = \cos\theta$, all the bounds can be reformulated with $\theta$ and $\phi$. 
Similarly, if we consider the density matrix of a polarized qubit as
\begin{align}
\rho=\frac{1}{2}\sum_{i=0}^{3} \frac{S_i}{S_0} \sigma_{i}, 
\end{align}
where $S_i (i=0,1,2,3)$ are Stokes parameters characterizing the polarization state and $\sigma_{i}$ are the Pauli matrices \cite{James}, then we can write $\langle \sigma_{i} \rangle = {S_i}/{S_0} $. Hence all terms in Eqs. (\ref{uncertainty-I}) to (\ref{uncertainty-VII}) can be reformulated with the Stokes parameters as 
\begin{align}
V=&\frac{1}{S_0^2} \left(S_1^2+S_2^2+S_3^2 \right) \; ,
\end{align}
\begin{align}
D=&\frac{1}{S_0^2} \left(S_1 S_2+S_2 S_3+S_3 S_1 \right) \; ,
\end{align}
\begin{align}
E=&\left| \frac{1}{S_0} \left(S_1+S_2+S_3 \right) \right|  \; ,
\end{align}
\begin{align}
H=&\left| \frac{S_1}{S_0}\right|+\left| \frac{S_2}{S_0}\right|+\left| \frac{S_3}{S_0}\right| \; ,
\end{align}
\begin{align}
L_{\pm}=&\sqrt{ 2-\left(\frac{S_1}{S_0} \pm \frac{S_2}{S_0} \right)^2 } \; ,
\end{align}
\begin{align}
M_{\pm}=&\sqrt{ 2-\left(\frac{S_2}{S_0} \pm \frac{S_3}{S_0} \right)^2 } \; ,
\end{align}
\begin{align}
N_{\pm}=&\sqrt{ 2-\left(\frac{S_3}{S_0} \pm \frac{S_1}{S_0} \right)^2 } \; .
\end{align}
These expressions are a heuristic for the further experimental study of classical light, i.e., it is an interesting question to explore the possible quantum behaviors of classical light with Stokes parameters.

\section{Conclusion}

In this work we report an experimental verification of three uncertainty relations for triple observables and four uncertainty relations for $N$ observables with single photon measurements. It is demonstrated that these uncertainty relations are valid for states of a spin-$1/2$ particle. Meanwhile, the experimental measurements exhibit the relative stringency of various uncertainty lower bounds, that is to say stronger or less stronger, within our experimental configurations. The present work enriches the experimental studies of multiple observables uncertainty relations, which 
are more general than pairwise ones in the discrete-dimensional Hilbert spaces. It is expected that the multi-observable uncertainty relation also needs to be tested with higher-dimensional systems, as the uncertainty relations are not only of fundamental importance to quantum theory, but also crucial to the quantum information technology like quantum precision measurement.

{\bf Acknowledgments}

\noindent This work was supported, in part, by the Ministry of Science and Technology of the People¡¯s Republic of China (2015CB856703); by the Strategic Priority Research Program of the Chinese Academy of Sciences, Grant No. XDB23030100; and by the National Natural Science Foundation of China (NSFC) under the Grants No. 11375200 and No. 11635009. S.M. Zangi is also supported, in part, by the CAS-TWAS fellowship.

\end{document}